\documentclass[epsfig]{aa}
\usepackage{epsfig,deluxe}
\begin{document}

\newcommand{\gsim}{\hbox{\rlap{$^>$}$_\sim$}}
  \thesaurus{06;  19.63.1}
%

\titlerunning{Supernova, Hypernova and Gamma Ray Bursts}
\title{Supernova, Hypernova and Gamma Ray Bursts\footnote{An expanded 
version of an invited talk presented at ``Young Supernova 
Remnants'', October 16-18, 2000
College Park, Maryland, USA, and based on work in collaboration
with A. De R\'ujula.}}

\author{Arnon Dar}
\institute{Physics Department and Space Research Institute, Technion,
              Haifa 32000, Israel}

\maketitle

\begin{abstract}

Recent observations suggest that gamma ray bursts (GRBs) and their
afterglows are produced by highly relativistic jets emitted in core
collapse supernova explosions (SNe). The result of the event, probably, is
not just a compact object plus a spherical ejecta: within days, a fraction
of the parent star falls back to produce a thick accretion disk around the
compact object. Instabilities in the disk induce sudden collapses with
ejection of highly relativistic ``cannonballs'' of plasma, similar to
those ejected by microquasars. The jet of cannonballs exit the supernova
shell/ejecta reheated by their collision with it, emitting highly
forward-collimated radiation which is Doppler shifted to $\gamma$-ray
energy. Each cannonball corresponds to an individual pulse in a GRB. They
decelerate by sweeping up the ionised interstellar matter in front of
them, part of which is accelerated to cosmic-ray energies and emits
synchrotron radiation: the afterglow. The Cannonball Model cannot predict
the timing sequence of these pulses, but it fares very well in describing
the total energy, energy spectrum, and time-dependence of the individual
$\gamma$-ray pulses and afterglows. It also predicts that GRB pulses
are accompanied by detectable short pulses of TeV neutrinos and sub TeV
$\gamma$-rays, that are much more energetic and begin and peak a little
earlier. 
 
\end{abstract} 


\section{Introduction}

Once upon a time, Gamma Ray Bursts (GRBs) constituted a sheer mystery,
whose unassailability was reflected in the scores of extremely different
ideas proposed to explain them. In spite of giant strides in the recent
observations ---the discovery of GRB afterglows (Costa et al. 1997; van
Paradijs et al. 1998), the discovery of the association of GRBs with
supernovae (Galama et al. 1998), and the measurements of the redshifts of
their host galaxies (Metzger et al. 1997)--- the origin of GRBs is still
an enigma. In the recent past, the generally accepted view has been that
GRBs are generated by synchrotron emission from fireballs, or firecones,
produced by collapse or merger of compact stars (Paczynski 1986; Goodman
et al. 1987; Meszaros and Rees 1992) by failed supernovae or collapsars
(Woosley 1993; Woosley and MacFadyen 1999; MacFadyen and Woosley 1999;
MacFadyen et al. 1999) or by hypernova explosions (Paczynski 1998). 

I have been asked to discuss hypernovae - hypothetical spherical fireballs
that are generated by gravitational, collapse of very massive stars and
generate GRBs. But ``I come to bury Caesar not to praise him'' (Shakespeare,
``Julius Caesar'' Act. III Sc.II): Various observations suggest that {\bf
most} GRBs are produced by highly collimated ultrarelativistic jets from
stellar collapse (Shaviv and Dar 1995; Dar 1998; Dar and Plaga 1999),
probably, from supernova explosions (Dar and Plaga 1999; Cen 1999; 
Woosley et al. 1999; Woosley and MacFadyen 1999;  MacFadyen and Woosley
1999; Dar and De R\'ujula 2000a;  Dar and De R\'ujula 2000b), and not in
spherical explosions that convert kinetic energy to GRBs with total
$\gamma$-ray energy in excess of $\rm 10^{54}~erg$. 

In my talk I will review briefly the evidence that GRBs are associated
with SNe. Then I will  review the Cannonball (CB) Model of GRBs that was
recently proposed  by Dar and De R\'ujula (2000a, 2000b) and explains how
GRBs are produced in SNe. Its success in describing the total energy,
energy spectrum, the time-dependence of the individual $\gamma$-ray pulses
in GRBs and the GRB afterglows will be demonstrated.

\section{The GRB--SNe association}

There is mounting evidence for an association of supernova (SN) explosions
and GRBs. The first example was GRB 980425 (Soffitta et al. 1998; Kippen
1998), within whose error circle SN1998bw was soon detected optically
(Galama et al. 1998) and at radio frequencies (Kulkarni et al. 1998a). The
chance probability for a spatial and temporal coincidence is less than
$10^{-4}$ (e.g. Galama et al. 1998), or much smaller if the revised
BeppoSAX position (e.g., Pian, 1999) is used in the estimate. The unusual
radio (Kulkarni et al. 1998a; Wieringa et al. 1999) and optical (Galama et
al. 1998; Iwamoto et al. 1998) properties of SN1998bw, which may have been
blended with the afterglow of GRB 980425, support this association. The
exceptionally small fluence and redshift of GRB 980425 make this event
very peculiar, a fact that we discuss in detail in section 7.

Evidence for a SN1998bw-like contribution to a GRB afterglow
(Dar 1999a) was first found by Bloom et al. (1999) for GRB 980326,
but the unknown redshift prevented a quantitative analysis.
The afterglow of GRB 970228 (located at redshift $\rm z=0.695$)
appears to be overtaken
by a light curve akin to that of SN1998bw (located at $\rm z_{bw}=0.0085$),
when properly scaled by their differing redshifts (Dar 1999b).
Let the energy flux density  of SN1998bw be $\rm F_{bw}[\nu,t]$.
For a similar supernova located at z:
\begin{eqnarray}
{\rm F[\nu,t] = }&&{\rm{1+z \over 1+z_{bw}}\;
{D_L^2(z_{bw})\over D_L^2(z)}}\, \times\nonumber \\
&&{\rm F_{bw}\left[\nu\,{1+z \over 1+z_{bw}},t\,
{1+z_{bw} \over 1+z}\right]\; A(\nu,z)}\, ,
\label{bw}
\end{eqnarray}
where $\rm A(\nu,z)$ is the extinction along the line of sight.
The SN--GRB association in the case of GRB 970228 was
reconfirmed by Reichart (1999) and by
Galama et al. (2000).  Evidence of  similar associations is found
for GRB 990712 (Hjorth et al. 1999; Sahu et al. 2000), GRB 980703 (Holland
2000) and GRB 000418: an example that we show
in Fig.~5. In the case of GRB 990510 the observational evidence
(Sokolov et al. 2000)
is marginal. For the remaining cases in Table I the observational data
preclude a conclusion, for one or more reasons: the late afterglow is not
measured; $\rm F_{bw}[\nu']$ is not known for large
$\rm \nu'\simeq \nu\,(1+z)$; the GRB's afterglow or the host galaxy are
much brighter than the SN. The case of GRB 970508, for
which the afterglow in the R band is brighter than a SN contribution
given by Eq.~(\ref{bw}), is shown in Fig.~6.

All in all, it is quite possible that a good fraction of GRBs are
associated with SNe, perhaps even {\it all} of the most frequent,
long-duration GRBs.
The converse statement ---that most SNe of certain
types are associated with GRBs--- appears at first sight to be untenable.
The rate of Type Ib/Ic/II SNe has been estimated from their observed
rate in the local Universe (e.g. Van den Bergh \& Tammann 1991)
and the star formation rate as function of redshift, to be 10 s$^{-1}$
in the observable Universe (Madau 1998).
The observed rate of GRBs is a mere 1000 y$^{-1}$.
Thus, very few of these SNe produce {\it visible} GRBs. But, if
the SN-associated GRBs were beamed within an angle $ \theta\sim
3.6\times 10^{-3}$,
only a fraction $\pi\,\theta^2/4\pi \sim 3\times 10^{-6}$ would be
visible, making the observed rates compatible and making possible a
rough one-to-one SN--GRB association
(or a ten-to-one association for $\theta\!\sim\! 1 \times 10^{-2}$).

\section{The Cannonball Model of GRBs}

\subsection{The engine}

The ejection of matter in a supernova (SN) explosion is not fully
understood. The known mechanisms for imparting the required kinetic energy
to the ejecta are inefficient: the theoretical understanding of
core-collapse SN events is still unsatisfying. It has been proposed (De
R\'ujula 1987; Dar and Plaga 1999; Cen 1999; Woosley and MacFadyen 1999;
Dar and De R\'ujula 2000a and references
therein) that the result of a SN event is not just a compact object plus
a spherical ejecta: a fraction of the parent star may be ejected, but another
fraction of its mass may fall back onto the newly born compact object. 
For vanishing angular momentum,
the free-fall time of a test-particle from a parent
stellar radius ${\rm R_\star}$ onto an object of mass ${\rm M_c}$ is:
\begin{eqnarray}
  {\rm t_{fall}}&&{\rm ={\pi\,\left[{R_\star^3\over 
8\,G\,M_c}\right]^{1/2} }}
\nonumber \\
&&{\rm \sim 1\; day\; \left[{R_\star\over 10^{12}\;cm}\right]^{3/2}\;
    \left[{1.4\;M_\odot\over M_c}\right]^{1/2}}\, .
\label{tfall}
\end{eqnarray}
The free-fall time may be shorter if the mass of the falling material is
not small relative to that of the compact object and if the specific
angular momentum is small, or much longer if the specific angular momentum
is considerably large, as it is in most stars. In both cases
it is quite natural to assume  
that infalling material with non-vanishing
angular momentum settles
into an orbiting disk, or a thick torus if its mass is comparable
to ${\rm M_c}$ and that, as observed
in other cases of significant accretion
onto a compact object (microquasars and active galactic nuclei)
in which the infalling material is processed in a series of
``catastrophic'' accretions,  jets of
relativistic CBs of plasma are ejected (e.g., Belloni et al. 1997; Mirabel 
and Rodriguez 1999a,b). Their
composition is assumed to be ``baryonic'', as it is in the jets of
SS 433, from which Doppler shifted Ly$_\alpha$ and K$_\alpha$ lines of 
various elements, such as Fe, Ni, Mg, Si, S and Ar, have been detected   
(Margon 1984; Kotani et al 1996), 
although the violence of the relativistic 
jetting-process
should in our case break most nuclei into their constituents.

The cannonball model of GRBs (Dar and De R\,ujula 2000a,b) is illustrated in 
Fig. 1. In brief, the CB model is the following. A sequence  
of highly relativistic cannonballs is emitted during 
a core-collapse SN. These cannonballs may be emitted right after the 
initial collapse (Dar and Plaga 1999; Cen 1999; Woosley and 
MacFadyen 1999; MacFadyen and Woosley 1999; MacFadyen et al.
1999) and hit a massive shell at a typical
radius of $\rm R_S\sim 3\times 10^{15}\, cm\, $ 
formed by strong wind emission in   
the Red Supergiant or variable Blue Giant presupernova phase 
(for evidence see e.g.,
Salamanca et al. 1998; Fassia et al. 2000). They may be emitted
in a second collapse (De R\'ujula 1987)  
at a time $\rm t_{fall}$ of ${\cal{O}}(1)$ day after
a SN core-collapse. By this time the SN outer shell, traveling at
a velocity $\rm v_S \sim c/10$ (see, e.g., Nakamura et al. 2000)
has moved to a distance:
\begin{equation}
\rm R_S=2.6 \times 10^{14} \;cm\;\left({t_{fall}\over 1\;d}\right)\;
\left({10\,v_S\over c}\right) .
\label{Rs}
\end{equation}
As it hits a massive shell, the CB slows down and heats up. Its radiation
is obscured by the shell up to a distance of order one radiation length
from the shell's outer surface. As this point is reached the emitted
radiation from the CB which continues to travel, expand and cool down,
becomes visible. This radiation, which is boosted and collimated by the
ultrarelativistic motion of the CB, and time contracted due to time 
aberration in the observer frame, appears as a single GRB pulse.  The 
observed
duration of a GRB pulse is its radiative cooling time after it becomes
visible. The total duration of a GRB with many pulses is the total
emission time of CBs with large Doppler factors by the central engine.  As
the mechanism producing relativistic jets in accretion is not well
understood, the CB model is unable to predict the timing sequence of the
successive GRB pulses but, the CB model is quite successful in describing
the total energy, energy spectrum and time-dependence {\it within single
GRB pulses} (Dar and De R\'ujula 2000b).

\subsection{Relativistic aberration, boosting and collimation} 

Let $\rm \gamma=1/\sqrt{1-\beta^2}={E_{CB}/(M_{CB}c^2)}$ be
the Lorentz factor
of a CB, that diminishes with time as the CB hits the SN shell
and as it subsequently plows through the interstellar medium.
Let $\rm t_{SN}$ be the
local time in the SN rest system, $\rm t_{CB}$ the time in the CB's
rest system and t the time of a stationary observer
viewing the CB at redshift z from an angle $\theta$
away from its direction of motion.
Let x be the distance traveled by the CB in the SN rest system.
The relations between the above timings are:
\begin{equation}
\rm dt= {(1+z)\, dt_{CB}\over \delta}={(1+z)\, dt_{SN}\over \delta\,\gamma}
={(1+z)\, dx \over \beta\,\, \gamma\, \delta\,  c}\,
\label{times}
\end{equation}
where the Doppler factor $\delta$ is:
\begin{equation}
\rm
\delta\equiv\rm{1\over\gamma\,(1-\beta\cos\theta)}
\simeq\rm {2\,\gamma\over (1+\theta^2\gamma^2)}\; ,
\label{doppler}
\end{equation}
and its approximate expression is valid for $\theta\ll 1$ and $\gamma\gg 1$,
the domain of interest here. In what follows we will set t=0 at the 
moment when the CB hits the shell.
Notice that for large $\gamma$ and $\theta\gamma\sim 1\, ,$
there is an enormous ``relativistic aberration'':
$\rm dt\sim dt_{SN}/\gamma\delta$ and the observer sees
a long CB story as a film in extremely fast motion.

The energy of the photons radiated by a CB
in its rest system, $\rm E^\gamma_{CB}$ and the photon
energy, E, measured by a cosmologically distant observer 
are related by:
\begin{equation}
\rm E={\delta\, E^\gamma_{CB}\over 1+z}\, ,
\label{energies}
\end{equation}
with $\delta$ as in Eq.(\ref{doppler}).
If $\rm E_{pulse}^{rest}$ is the CB total emitted 
radiation in its rest frame, 
an observer at a luminosity distance $\rm D_L(z)$ from the CB that is
viewing it at an angle $\theta<<1$ from its direction of motion would measure
a  ``total'' (time- and energy-integrated) fluence per unit area:
\begin{equation}
\rm {dF\over d\Omega}\simeq {(1+z)\, E_{pulse}^{rest}
\over 4\,\pi\,D_L^2}\;\delta^3\; .
\label{dfdomega}
\end{equation}
Only if traveling with a large Lorentz factor $\gamma > 10^2$ at a small
angle $\theta\sim 1/\gamma $ relative to the line of sight, will a CB 
at a cosmological distance be visible.

\subsection{Jet energy and CB mass}

Let ``jet'' stand for the ensemble of CBs emitted in one direction in a SN
event. If a momentum imbalance between the opposite-direction jets is
responsible for the large peculiar velocities ${\rm v_{NS}\approx 450\pm
90~ km~s^{-1}}$ (Lyne and Lorimer 1994) of neutron stars born in SNe, the
jet kinetic energy $\rm E_{jet}$ must be, as assumed in the CB model for 
the GRB
engine, larger than $\sim 10^{52}$ erg (e.g. Dar and Plaga 1999). The
jet-emitting process may be ``up-down'' symmetric to a very good
approximation, in which case the jet energies may be much bigger. There is
evidence that in the accretion of matter by black holes in quasars
(Celotti at al. 1997; Ghisellini 2000) and microquasars (Mirabel and
Rodriguez 1999a,b) the
efficiency for the conversion of gravitational binding energy into jet
energy is surprisingly large.
If in the production of CBs the central compact object
in a SN ingurgitates several solar masses, it is not
out of the question that $\rm E_{jet}$ be as large as
$\rm M_\odot c^2\sim 1.8\times 10^{54}$ erg. 
A compromise value, $10^{53}$ ergs, as the reference
jet energy was adopted in the CB model of GRBs.

Average GRBs have some five to ten significant pulses, so that the 
fraction f of the jet energy carried by a single CB
may typically be 1/5 or 1/10. A value $\rm E_{CB}=10^{52}$ erg
was adopted as a reference value. For this value, the CB's mass is
comparable to an Earth mass:
${\rm M_{CB}\sim 1.8\, M_\otimes (10^3/\gamma)}$, for a Lorentz factor
of $\rm\gamma={\cal{O}}(10^3)$, that was found by fitting the GRB 
properties (see later).

\subsection{The making of the GRB}

The general properties of GRB pulses in the CB model are not sensitive to the
complex details of the CB's collision with the shell. They can be
estimated from the overall energetics and approximate treatment of the
cooling of the CB by radiation and expansion as it reaches the 
transparent outskirts of the shell. 

The density profile of the outer layers of an SN shell as a function
of the distance x to the SN center can be inferred from the photometry,
spectroscopy and evolution of the SN emissions (see e.g. Nakamura et al. 2000 
and references 
therein). The observations can be fit by a power law,
$\rm x^{-n}$, with $\rm n \sim 4\; to\, 8$. 
The results are sensitive to
this density profile only in the outer region where the SN shell
becomes transparent (and the measurements are made), so that one
can adopt the same profile at all $\rm x>R_S$:
\begin{equation}
\rm \rho(x)=\rm\rho(R_S)\,\Theta(x-R_S)\,\left[{R_S\over x}\right]^n\, .
\label{profile}
\end{equation}
The SN-shell grammage still in front of a CB located at x is:
\begin{equation}
\rm X_S(x)=\int_x^\infty \, \rho(y)\,dy=
{M_S\over 4\,\pi\, R_S^2}\; \left[{R_S\over x}\right]^{n-1}\, ,
\label{SNgram}
\end{equation}
where $\rm M_S$ and $\rm R_S$ are the total mass and radius, 
respectively, of the shell.
For photons in the MeV domain the attenuation length is similar, within
a factor 2, in all elements from H to Fe (Groom et al., 2000), and can be 
roughly approximated by:
\begin{equation} 
\rm X_\gamma(E)\sim 1.0\,(E/keV)^{0.33}\; g\, cm^{-2}\; .
\label{Xgamma}
\end{equation}
The value of $\rm X_\gamma(E)$ in the $\rm E=10$ keV to 1 MeV domain
(2.1 to 9.8 gr/cm$^2$) is close to the attenuation length in a hydrogenic
plasma ($\rm X_\gamma^{ion}\simeq m_p/\sigma_{_T}\simeq 2.6$ gr/cm$^2$,
with $\rm m_p$ the proton's mass and
$\rm \sigma_{_T}\simeq 0.65\times 10^{-24}$ cm$^2$ the Thomson 
cross-section). Therefore, it makes little difference in practice whether
or not we take into account that the SN-shell material reached
by the CB may be ionized by its previously emitted radiation.
The position $\rm X-{tp}$ at which the SN shell becomes
(one-radiation-length) transparent is then given by :
\begin{equation}
\rm x_{tp}(E) = R_S\;\left[{M_S\over 4\,\pi\, R_S^2}\;
{1\over X_\gamma(E)}\right]^{1\over n-1}\propto E^{-0.33/(n-1)}\; ,
\label{SNtransparent}
\end{equation}
whose energy dependence is extremely weak.
Blue-shifted to the SN rest-system, as in Eq.(\ref{energies}),
GRB photons have energies
in the MeV range. 

In its rest frame, the front surface of the CB is bombarded by the nuclei
of the shell, which have an
energy $\rm m_p \,c^2\,\gamma\sim$ 1 TeV per nucleon,
roughly 1/3 of which (from $\pi^0\to\gamma\gamma$ decays)
is converted into these
$\gamma$-rays within $\rm X_p\approx m_p/\sigma_{in}(pp)\approx 50\, g\,
cm^{-2}$, where $\rm \sigma_{in}(pp)$ is the nucleon-nucleon
inelastic cross section.
These high energy photons initiate electromagnetic cascades that
eventually convert their energy to thermal energy within the CB.
The radiation length of high energy $\gamma$'s in hydrogenic plasma,
dominated by $\rm e^+\,e^-$ pair production, is $\rm X_{\gamma e}
\simeq 63$ g cm$^{-2}$,
comparable to $\rm X_p$. The radiation length of thermalized
photons in a hydrogenic plasma is
$\rm X_\gamma^{ion}\approx m_p/\sigma_{_T}\approx 2.6$ g cm$^{-2}$.

Assume that the quasi-thermal emission rate from
the CB, within $\rm X_\gamma^{ion}$ from its
surface, is in dynamical equilibrium with the fraction of energy deposited
by the CB's collision with the shell in that outer layer.
The temperature of the CB's front is then
roughly given by:
\begin{equation}
\rm T(x)\simeq \left[{(n\! -\! 1)\,X_\gamma\, m_p\, c^3\, \gamma^2\,
              \sigma_{in}(pp)  \over
              6\,\sigma\, x_{tp}\,X_{\gamma e}
\, \sigma_{_T}^2}\right ]^{1\over 4}
              \left[ {x\over x_{tp}}\right]^{-{n\over 4}}\!\!\! ,
\label{temperature}
\end{equation}
Remarkably, only the Lorentz factor of the CBs when they exit the  shell, 
but neither their mass
nor their energy,  appear in the above expression,
except for the fact that, for the result to be correct, they must be
large enough for the CB to pierce the shell and remain relativistic.

The CB temperature at $\rm t_{tp}$ is not sensitive to the exact
value of n, unlike its time dependence.  
For $\rm n=8$ and $\rm M_s=10\, M_\cdot$ 
the value of 
$\rm x_{tp}$ is $\rm \approx 3 \,R_S$, and for t close
to $\rm t_{tp}$ or later:
\begin{equation}
\rm T(t)\simeq 0.16\, keV\,
        \left[ {t_{tp}\over t}\right]^{2}\,\left[ {\gamma(t)\over
10^3}\right]^{1\over 2}\; .
\label{newtemp}
\end{equation}
This estimate is valid as long as the surface temperature 
of the optically thick CB is higher than the internal temperature of the 
CB that under the assumption of isentropic expansion decreases with time
only like $\rm T_{CB}\sim 1/R_{CB}\sim 1/t)$.

The observed energy and time dependence of the photon intensity 
(photon number per unit area, N) 
of a single pulse  in a GRB at an angle $\theta$ relative to the CB's
motion is  predicted to be:
\begin{eqnarray}
&&\rm {dN\over dE\,dt}\equiv {1+z\over 4\,\pi\,D_L^2}\;
\delta^2\,  {dn_\gamma\over dE\,dt}\, ,\\
&&
\rm {dn_\gamma\over dE\,dt}\simeq {2\,\pi\,\sigma\over \zeta(3)}\;
{\left[R_{CB}[t]\;E\,(1+z)/\delta\right]^2\; Abs(E,t)\over 
Exp\left\{E\,(1+z)/(\delta\, T[t])\right\}-1}\; ,
\label{boostthermal}
\end{eqnarray}
with $\rm R_{CB}[t]\simeq c\,t/\sqrt{3}\gamma $ and $\rm T[t]$ as in 
Eq.(\ref{temperature}), and where
\begin{equation}
\rm Abs(E,t)=
Exp\left[-{X_S(x[t])\over X_\gamma(E\,(1+z))}\right]
\label{attenuation}
\end{equation}
is the attenuation of the flux in the shell.

The mean photon energy of a black body radiation 
is approximately 2.7 T. Thus, around
peak energy flux, the mean energy of the observed photons is
\begin{equation}
\rm <E_\gamma> \simeq {0.45 \over  1+z}\,\left[{\delta\over 10^3} \right]
\, MeV\; .
\label{meane}
\end{equation}
It yields $\rm <E_\gamma>\simeq 0.23\, MeV$ for $\delta\simeq 10^3$ and z=1
--the mean redshift of the GRBs listed in Table I--, in good agreement with
that measured in GRBs (Preece 2000). 

For $\rm n=4$
the temperature decrease approximately
as 1/t. For n $>4$ it diminishes faster than 1/t and for $\rm n=8$
it decreases faster than
$\rm 1/t^2$, the ``faster'' being due, in both cases, to the effect
of a decreasing $\rm \gamma(t)$. For  $\rm T\sim 1/t^2$ behaviour,
the pulse width narrows with time like  $\rm \sim E^{-0.5}$
in agreement with the analysis of
Fenimore et al.~(1995) who
found, from a large sample of GRB pulses, that it narrows like
$t \propto E^{-0.46}$.

The total radiated energy, in the CB rest frame, is roughly the thermal
energy deposition within one radiation length from its
front surface. After attenuation in the SN shell, it reduces to:
\begin{equation}
\rm  E_{pulse}^{rest}\approx  {\sigma_{in}(pp)\, \pi\,
[R_{CB}^{tp}]^2\,\bar X_\gamma\, m_p\,c^2\, \gamma(t)
                      \over 3\,X_{\gamma e}\, \sigma_{_T}^2}\,,
\label{newenergy}
\end{equation}
where  $\rm  \bar X_\gamma$ is the radiation length
in the obscuring shell averaged over the black body spectrum.
For a typical $\gamma$-ray
peak energy of $\rm E_p\sim 1\, MeV$ in the SN rest frame,
$\rm \bar X_\gamma\simeq 10\ g\, cm^{-2}\,. $ Consequently,
the CB's radius at transparency is
$\rm R_{CB}^{tp}=4\times 10^{11}$ cm and
$\rm  E_{pulse}^{rest}\sim 3\times 10^{45}\, erg$,
for $\rm\gamma(t)\sim 10^3$.  

This is consistent with the estimated GRB energies in Table I provided 
that the typical Doppler factors of GRBs are in the range
$300\leq \delta\leq 1000$.

\section{Some simplifications and approximate predictions}

A CB, in
its rest system, is subject to a flux of high energy nuclei and electrons.
While the electrons are being thermalized, they contribute a
nonthermal high-energy tail of photons emitted via the ``free-free''
process. Such a power-law tail in an otherwise
approximately-thermal emission is observed from
young supernova remnants (see, e.g., Dyer et al. 2000) and clusters of
galaxies (e.g., Fusco-Femiano et al. 1999; Rephaeli et al., 1999;
Fusco-Femiano et al., 2000), both of which are systems
wherein a dilute plasma at a temperature of
$\cal{O}$(1 keV) is exposed to a flux of high energy cosmic rays.
Thus, the CB emission can be modeled as a black body spectrum 
with a nonthermal power-law tail. 
If, for the sake simplicity, 
the energy spectrum of the surface radiation  
from a CB is approximated by a thermal black body radiation  then, 
Eqs. (\ref{boostthermal}--\ref{attenuation}) yield    , 
\begin{equation}
\rm {dN\over dE\,dt}\propto
{(E\,t)^2\over Exp\{E\,t/H\}-1}\,
Exp \left\{-\left[ {t_{tp}/ t}\right]^{n-1}\right\}
\;\Theta[t]\; .
\label{simple}
\end{equation}
The total photon intensity and energy flux are, in this 
approximation:
\begin{equation}
\rm {dN\over\,dt}\propto
\Theta[t]\; {t_{tp}\over t}\, Exp 
\left\{-\left[ {t_{tp}/ t}\right]^{n-1}\right\}\,, 
\label{simple2}
\end{equation}
\begin{equation}
\rm {F_E(t)}\propto
\Theta[t]\; \left[{t_{tp}\over t}\right]^2\,
Exp \left\{-\left[ {t_{tp}/ t}\right]^{n-1}\right\}\,.
\label{simple3}
\end{equation}
Let the peak $\gamma$-ray 
energy at a fixed time during a GRB pulse be defined as
$\rm E^\gamma_p(t) \equiv max\,[ E^2\,dI_\gamma/dE\, dt]$.
Its value is   $\rm E^\gamma_p(t)\simeq 3.92\,\delta\,T[t]/(1+z) $, so that,
for t near or after $\rm t_{tp}$:
\begin{equation}
\rm E^\gamma_p(t) \simeq E^\gamma_p(t_{tp})
\;\Theta[t]\; {t_{tp}\over t}\,.
 \label{simple4}
\end{equation}

The total ``isotropic'' energy of a GRB pulse 
-- inferred from its observed 
fluence assuming an isotropic emission-- can be deduced from
Eq.~(\ref{dfdomega}), to be:
\begin{equation}
\rm   E_{iso}=
{4\,\pi\,D_L^2\, F \over 1+z}\simeq E_{pulse}^{rest}\, \delta^3\,.
\label{eisotropic}
\end{equation}

If CBs were  ``standard candles'' with fixed mass, energy
and velocity of expansion,
and if all SN shells had the same 
mass, radius and density distribution, all differences between 
GRB pulses would result from their different distances and angles
of observation. For such standard candles 
it follows from Eqs.(\ref{times}-\ref{energies},\ref{simple2}, 
\ref{simple3}) that the observed
durations (half widths at half maximum)  of the photon intensity and 
of the energy flux density ($\rm \Delta t_I$ and
 $\rm \Delta t_F$),
their peak values  ($\rm N_p$ and $\rm F_p)$, and the peak energy 
($\rm E^\gamma_p$) in a single GRB pulse are 
roughly correlated to the total ``observed'' isotropic energy 
($\rm E_{iso}$) as follows:
\begin{equation}
\rm \Delta t_I\propto (1+z)\, [E_{iso}]^{-1/3}\,,  
\label{twidthi}
\end{equation}  
\begin{equation}
\rm \Delta t_F\propto (1+z)\, [E_{iso}]^{-1/3},  
\label{twidthf}
\end{equation}  
\begin{equation}
\rm N_p\propto E_{iso},  
\label{Ipeak}
\end{equation}  
\begin{equation}
\rm F_p\propto[ E_{iso}]^{4/3}\, (1+z)^{-1}\, , 
\label{Lpeak}
\end{equation}  
\begin{equation}
\rm E^\gamma_p\propto [E_{iso}]^{1/3}\,(1+z)^{-1}\, .  
\label{Epeak}
\end{equation}  
These approximate correlations can be tested using the sample of 15 GRBs with
known redshifts. 
Because of the strong dependence of the CB pulses 
on the Doppler factor and their much weaker dependence on the 
other parameters, they may be approximately satisfied
(see, e.g. Plaga 2000) in spite of the fact 
that CBs and SN shells are likely to be sufficiently varied
not to result in standard candles.

\section{Predictions of the Cannonball Model}

Some common properties of GRB pulses
(for detailed light curves see Kippen 2000; Mallozzi 2000) are observed
to be:
\begin{itemize}
\item{(a)} The GRB fluences, integrated in energy and time,
lie within one or two orders of magnitude above or below
10$^{-5}$ erg/cm$^2$ (see, e.g., Paciesas et al. 1999).
\item{(b)} Individual pulses are narrower in time, the higher the
energy interval of their individual photons
(see, e.g., Fenimore et al. 1995).
\item{(c)} Individual pulses rise and peak at earlier time, the higher the
energy interval of their individual photons
(see, e.g., Norris et al. 1999; Wu and Fenimore 2000)
\item{(d)} Individual pulses have smaller photon energies, the
later the time-interval of observation (see, e.g., Preece et al. 1998) .
\item{(e)} The energy spectrum of GRBs, or of their individual
pulses, if plotted as $\rm E^2\,dN/dE$, rises with energy as $\rm E^\alpha$,
with $\alpha \sim 1$,  has a broad peak at $\rm E\sim 0.1$ to 1 MeV, and
decreases thereafter (see, e.g., Preece et al. 2000).
\item{(f)} Most GRBs consist of pulses whose time-behaviour is
a fast rise followed
by an approximately exponential decay: a ``FRED'' shape. Some
GRBs have non-FRED, roughly  time-symmetric pulses
(see e.g., Fenimore et al. 1995 and references therein)
The overwhelming majority of GRBs are either made of FRED or non-FRED
pulses: there are no GRBs with mixed pulse-shapes.
\end{itemize}

All of the above items are properties of the CB model, as was shown in Dar
and De R\'ujula 2000b. Figs. 2-4, taken from this reference, demonstrate
the success of the CB model in explaining the temporal structure of GRBs,
the spectrum of individual pulses and their temporal evolution. 

\section{GRB afterglows}

Far from their parent SNe, the CBs are slowed down by the interstellar
medium (ISM) they sweep, which has been previously ionized by the
forward-beamed CB radiation (traveling essentially at $\rm v=c$, the CB
is ``catching up'' with this radiation, so that the ISM has no time to
recombine). As in the jets and lobes of quasars, a fraction of the
swept-up ionized particles are ``Fermi accelerated'' to cosmic-ray
energies and confined to the CB by its turbulent magnetic field,
maintained by the same confined cosmic rays (Dar and Plaga
1999). The bremsstrahlung and synchrotron  emissions from the high energy 
electrons in the CB, boosted by its
relativistic bulk motion, produce afterglows in all bands
between radio and X-rays, collimated within an angle $\sim 1/\gamma(t)$,
that widens as the ISM decelerates the CB. 

A CB of roughly constant cross section, moving in a
previously ionised ISM of roughly constant density, 
slows down according to 
${\rm d\gamma/dx=-\gamma^2/x_{0}}$, with ${\rm x_{0}=M_{CB}/
(\pi\, R_{CB}^2\, n\, m_p})$ and n the number density along
the CB trajectory. 
For $\gamma^2\gg 1$,
the relation between the length of
travel dx and the (red-shifted, relativistically aberrant) time of 
an observer at a small angle $\theta$ is 
${\rm dx=[2\, c\, \gamma^2/(1+\theta^2\,\gamma^2)]\,[dt/(1+z)]}$.
Inserting this into $\rm d\gamma/dx$ and integrating, one obtains:
\begin{equation}
{\rm {1+3\,\theta^2\gamma^2\over 3\,\gamma^3}=
{1+3\,\theta^2\gamma_0^2\over 3\,\gamma_0^3}+
{2\,c\, t\over (1+z)\, x_{0}}}\; ,
\label{gamoft}
\end{equation}
where $\gamma_0$ is the Lorentz factor of the CB as it exits
the SN shell.
The real root $\rm \gamma=\gamma(t)$ of the cubic Eq.~(\ref{gamoft})
describes the CB slowdown with observer's time. 

The radiation emitted by a CB in its rest system
(bremsstrahlung, synchrotron, Compton-boosted 
self-synchrotron), 
is boosted and collimated 
by the CB's motion, and its time-dependence is
modified by  the observer's 
time flowing  $(1+z)/\delta$ times
faster than in the CB's rest system. 
For $\gamma \gg 1$,
an observer at small $\theta$ sees an energy flux density:
\begin{equation}
\rm F[\nu] \sim  \delta^3\, F_0(\nu\,[1+z]/\delta)\, A(\nu,z)\, , 
\label{sync}
\end{equation}    
where $\rm F_0(\nu_0)$ is the CB emission in its rest frame, 
$\rm \delta(t)$ is given by Eq.~(\ref{doppler}) with 
$\rm\gamma=\gamma(t)$ as in Eqs.~(\ref{gamoft}), and $\rm A(\nu,z)$ 
an eventual absorption dimming. 

Neglecting energy deposition through collision of the CB with the ISM 
during the afterglow regime
and energy losses due to expansion and radiation, the CB's afterglow  is
dominated by a steady 
electron synchrotron radiation from the 
magnetic field in the CB 
(valid only for slowly expanding/cooling CBs in
a low density ISM and 
for frequencies where the attenuation length due to electron free--free
transitions exceeds its radius).  The spectral shape of synchrotron
emission in the CB rest frame is ${\rm F_0\sim \nu_0^{-\alpha}}$, with
${\rm \alpha=(p-1)/2}$ and p the spectral index of the electrons. For
equilibrium between Fermi acceleration and synchrotron and Compton
cooling, $\rm p \approx 3.2$ and $\alpha\approx 1.1$, while for small
cooling rates, $\rm p \approx 2.2$ and $\alpha \approx 0.6$ (Dar and De
R\'ujula 2000c), or $\rm p\simeq1.2$ and $\alpha\simeq 0.1$ if Coulomb
losses dominate.  At very low radio frequencies self-absorption becomes
important and $\alpha \approx -1/3~(2.1)$ for optically thin (thick) CBs.

Since $\rm \gamma(t)$, as in Eq.~(\ref{gamoft}), is a decreasing
function of time, the
afterglow described by Eq.~(\ref{sync})
may have a very interesting behaviour.
An observer may initially be outside the beaming cone: 
$\theta^2\gamma^2\! >\! 1$, as we shall argue to be the case for
GRB 980425, for which we estimate $\gamma_0^2\,\theta^2\!\sim\! 200$ 
(other relatively dim GRBs in Table I, such as 970228 and 970508, 
may also be of this type). 
The observed afterglow would then initially rise with time. 
As $\gamma$ decreases, the cone
broadens, and around $\rm t=t_p$ when  $\gamma\theta\!\sim\! 1$ (when 
the observer enters the `beaming cone' of the CB)  
the afterglow  peaks and then begins to decline. 
Beyond the peak, when $\gamma^2\theta^2\gg 1$ and where 
$\rm \gamma\sim t^{-1/3}$, the afterglow declines like 
\begin{equation}
\rm F[\nu] \sim  F[\nu,\,t_p]\, 
\left [{(t_p/t)^{1/3}
\over 1+(t_p/t)^{2/3}}\right]^{3+\alpha+\beta}
\, A(\nu,z)\, ,
\label{synclate}
\end{equation}    
where $\rm F_0 \sim \nu^{-\alpha}\, t^{-\beta} $ in the CB rest frame.    
For steady emission, i.e., for small emission and energy deposition rates,  
$\beta=0$. For an emission rate which is in equilibrium with 
the energy deposition rate in the CB rest 
frame by the incident ISM particles  (that is proportional to 
$\gamma^2$), $\beta=2/3$. Eq.~(\ref{synclate}) describes well the late
afterglows of most GRBs, including  their radio afterglows (if one 
correct it for absorption in the CB and along the line of sight).

This is demonstrated in Fig. 5  for GRB 000418 where 
its late afterglow in the R band as predicted by
Eqs.(\ref{gamoft},\ref{synclate}) is compared  with the observations 
that were compiled in  Klose et al. 2000.
The normalization and $\rm t_p$  were adjusted to
fit the data. A contribution from a SN1998bw-like SN placed at the GRB
redshift, z=1.11854, as in Eq.(\ref{bw}) with Galactic extinction ${\rm
A_R=0.09}$ magnitudes was added to the CB light curve.

Eqs.~(\ref{gamoft},\ref{sync}) 
may explain the puzzling initial rise 
of the optical afterglow of GRBs 970228 and 970508,
as well as the second peak around $\rm t_p\sim 3$--$\rm 4\times 10^6\, s$ 
in the unresolved radio emission 
from SN1998bw/GRB 980425  (Kulkarni et al. 1998a; Frail et al. 1999),
if it corresponds to the GRBs afterglow. This is demonstrated in 
Fig. 6. where the afterglow of GRB 970508 in the R band is  compared 
with Eqs.~(\ref{gamoft},\ref{sync}), for the measured index
$\alpha=1.1$. The adjusted parameters are
the height, the time of the afterglow's peak and the product 
$\theta\gamma_0$.
The figure is for a single CB; with a few of them at chosen times
and relative fluxes,
it would be easy to explain the early ``warning shots'' at $\rm t<1$ day
and the abrupt rise at $\rm t=1$ to 2 days. At $\rm t\gg t_p$, however,
they would add up to a single curve like the one shown in the figure.

When the CB  enters the Sedov--Taylor phase
its radius increases as ${\rm t^{2/5}}.$
The Lorentz factor of electrons decreases like ${\rm  t^{-6/5}}$.
In equipartition, the magnetic field decreases like ${\rm t^{-3/5}}$.  
Wijers et al. (1997) have shown that these facts lead to an afterglow
decline ${\rm F_\nu\sim t^{-(15\alpha-3)/5}\sim  t^{-2.7}}$
for  $\alpha\sim 1.1$  CB at rest,    
as was observed for the late-time afterglows of
some GRBs. Note, however, that if the CB enters the 
Sedov-Taylor phase in flight, Eq.~(\ref{sync}) 
predicts  $\rm F_\nu \sim t^{-{(50\alpha+6) / 15}}$, i.e., 
$\rm F_\nu\sim t^{-2.4}$ for $\alpha\sim 0.6$ which changes to 
$\rm F_\nu\sim t^{-4}$ for $\alpha \sim 1.1$.

\section{GRB 980425: a special case}

In the list of Table I, GRB 980425 stands out in two ---apparently 
contradictory---
ways: it is, by far, the closest ($\rm z=0.0085$, $\rm D_L=39$ Mpc) 
and it has, by far, the smallest implied spherical energy: ${\rm 
8.1\times 10^{47}}$ erg,  4 to 6 orders of magnitude smaller than that 
of other GRBs. 
However,if GRB 980425 was not abnormal, Eq.~(\ref{eisotropic}) 
tells us that the peculiarities of of GRB 980425 can be understood
if its source was ``fired''  from SN1998bw with a bulk-motion Lorentz factor
$\gamma \sim 1000$, at an angle $\theta\approx 15/\gamma$ relative to our
line of sight. Then, for $\theta\ll 1$ and $\gamma\gg 1$,  
its projected sky velocity  
\begin{equation}
{\rm v_T^+\approx {2\,\gamma^2\,\theta\over (1+\gamma^2\theta^2)}\;c}\, ,
\label{superlum}
\end{equation}   
yields superluminal velocities (Rees 1966)
${\rm v_T\approx 2\,c/\theta}$
for $\gamma^2\theta^2\gg 1$, 
and ${\rm v_T\approx 2\,c\,\gamma^2\,\theta}$,
for $\gamma^2\theta^2\ll 1$, 
provided ${\rm 2\,\gamma^2\,\theta\! >\! 1}$.
Its transverse superluminal
displacement, $\rm D_T$, from the SN position can be obtained by 
time-integrating
$\rm v_T$, as in Eq.~(\ref{superlum}), using $\rm \gamma(t)$ as in
Eq.~(\ref{gamoft}). The result can be reproduced, to better than 10\%
accuracy, by using the approximation $\rm v_T\sim 2 \gamma^2\,\theta\,c$,
valid for $\gamma\! <\! 1/\theta$ (or $\rm t\! >\! t_p$, the afterglow's
peaktime):  
\begin{equation} 
{\rm D_T\simeq{2\,c\, t_p\over \theta} \left[{t\over t_p}\right]^{1/3}}.  
\end{equation} 
At present ($\rm t\!\sim\! 850\, d$),
the displacement of the GRB from the initial SN/GRB position is $\rm
D_T\!\sim\! 20\,(\gamma_0/10^3)$ pc, corresponding to an angular
displacement $\Delta\alpha\!\sim\!100\,(\gamma_0/10^3)$ mas, for $\rm
z=0.0085$, $\rm t_p\!\sim\! 3.5\times 10^6$ s.
 
The two sources, now a few tens of mas away, may still be resolved by HST. 
In Fig.~7 we show our prediction for the late-time V-band light curve of
SN1998bw/GRB 980425. The SN curve is a  fit by Sollerman 
et al. (2000) for energy deposition by $\rm ^{56}Co$ decay  
in an optically thin SN shell.  The GRB light curve is our predicted 
afterglow for GRB 9980425, as given by Eqs.~(\ref{gamoft},\ref{sync}),
and constrained to peak at the position of the second peak in
the radio observations (Kulkarni et al. 1998a; Frail et al. 1999).
The fitted normalization  
is approximately that of the mean afterglow of the GRBs in Table I,
suppressed by the same factor as its $\gamma$-ray fluence relative to 
the mean $\gamma$-ray fluence in Table I.  
The joint system has at present (day $\sim\! 850$) an extrapolated   
magnitude $\rm V\sim 26$ (Fynbo et al. 2000). 
It  can still be resolved  from its
host galaxy ESO-184-G82 by HST and, perhaps, by VLT in good seeing
conditions.  An extrapolation of the V-band late time
curve of SNR1998bw (Sollerman et al. 2000) suggests
that the present magnitude of the SN is $\rm V\sim 28$ , which is near
the detection limit of HST and is dimming  
much faster than $\rm ^{56}Co$-decay would imply.

In the GHz radio band, the system has been last observed  in February 1999
by ATCA (Australia's Telescope Compact Array) to have an 
${\cal{O}}(1)$ mJy
flux density (Frail et al. 2000) and to have approached a power-law time decline
with a power-law index $-1.47$. If a Sedov-Taylor break in the radio afterglow 
of GRB 980425 has not occurred yet, its
spectral density may still be strong enough to determine its position with
ATCA and VLBI to better than mas precision. If the second peak is the GRB's
afterglow, the two radio centroids should now be separated by
$\sim\!\gamma_0/10$ mas (or by $\sim\!\gamma_0/20$ mas, if the first peak
is the afterglow).  A refined location from a reanalysis of the early ATCA
observations (Kulkarni et al 1998a; Frail et al. 1999) of the initial SN
and of the late afterglow may also reveal a superluminal displacement.

If the GRB  afterglow  has entered the late
fast-decline phase seen in some GRB afterglows and in quasar and
microquasar ``afterglows'' from jetted ejections (observed power-law index
$-2.7 \pm 0.3$),  a further delay in follow-up observations can make it
very difficult or impossible to detect and resolve the GRB/SNR radio image
into its two  predicted images.  

A GRB as close as GRB 980425 (z = 0.0085)  should occur only once every
$\sim\!10$ years and its associated supernova may be only occasionally
observed. For typical GRBs ($\rm z\!\sim\! 1$) there is no hope of 
resolving them with HST into two separate SN and GRB images. Resolving
them with VLBI would also be arduous. For these reasons, we
exhort interested observers to consider immediate high-resolution optical
(STIS)  and radio (ATCA and VLBI) follow-up observations of SN1998bw and
the afterglow of GRB980425. 

\section{X-ray lines in GRB afterglows}

Lines in the X-ray afterglow 
of GRBs have been detected in four GRBs (GRB 970508, Piro et al. 1999;
GRB 970828, Yoshida et al 1999; GRB 991216, Piro et al. 2000;
GRB 000214, Antonelli et al. 2000). Their energies are listed 
in Table II. They were interpreted as iron lines emitted from
a large mass of iron that was photoionized by the GRB. However,
this interpretation raises many questions (see, e.g., Vietri 2000).   

\subsection{Origin of the X-ray lines}

The CB model offers an alternative interpretation for the origin of
the X-ray lines in the ``early'' GRB afterglow - hydrogen recombination 
lines which are Doppler shifted to X-ray energies by the CB
relativistic motion:
 
As long as the CB is opaque to its internal radiation, it 
expands with the relativistic 
speed of sound, $\rm c\sqrt{3}$ in its rest frame, 
and cools with ${\rm
T_{CB}\sim 1/R_{CB}\sim 1/t}$. 
When 
${\rm R_{CB}\simeq [3\,M_{CB}\,\sigma_T/(4\, \pi\,
m_p)]^{1/2}}$, where $\rm \sigma_T\simeq 0.65\times 10^{-24}$ cm$^2$ is
the Thomson cross section, it becomes
optically thin and its internal radiation escapes. 
This end to the CB's $\gamma$-ray pulse
takes place at $\rm t\simeq R_{CB}/\sqrt{3}\,c\,\delta\simeq 2\times
(10^3/\delta)\,s$ in the observer frame. 
Due to the escape of its internal radiation, 
its internal pressure drops and its expansion rate
is slowed down by sweeping up the ISM.
During this phase it cools mainly
by emission of bremsstrahlung and synchrotron radiation.
Its ionization state is described by the Saha
equation
\begin{equation}
{\rm {x^2 \over 1-x} = {(2\, \pi\, m_e\, c^2\, k\,T)^{3/2}\over
                     n\, h^3\, c^3}\, e^{-\chi/k\,T}}\,
\end{equation}
with  ${\rm x=n_e/n}\,,$  n being the the baryon density in the shell,
$\chi=13.6$ eV is the hydrogen binding energy
T is the plasma temperature in K and
$\rm n_e$ is the density of free electrons
in $\rm cm^{-3}\,.$ 
When the electron temperature
in the CB approaches 5000 K, electrons begin to recombine with protons 
into hydrogen.  The exponential term in the Saha equation
confines this recombination phase of the CB to a temperature around 4500K 
(for CBs with $\rm  10^5\, cm^{-3} <n_e<10^6\, cm^{-3}\,,$ as we shall 
estimate later).   The  recombination produces strong emission    
of $\rm Ly_{\alpha}$ line (and, perhaps, a recombination edge 
above the $\rm Ly_{\infty}$ line)  which is Doppler shifted 
by the CB motion to X-ray energy in the observer frame:
\begin{equation} 
\rm E_\alpha\simeq {10.2\over(1+z)}\,\left[ {10^3\over \delta}\right]\ 
keV\, ,   
\end{equation}   
\begin{equation} 
\rm E_{edge}\simeq {13.6\over(1+z)}\,\left[{10^3\over \delta}\right]\, 
keV\,.   
\end{equation}   
The total number of these recombination photons is approximately
equal to the baryonic number of the CB, 
$\rm N_b\simeq E_{CB}/m_p\,c^2\,\gamma(0)\simeq 6.7\times 10^{51}\, 
(E_{CB}/10^{52}\, erg)\, (10^3/\gamma(0))\,.$
Thus, the photon fluence of these
lines (line-fluence) at a luminosity distance $\rm D_L$ is
\begin{equation}
\rm N_{lines}\simeq N_b\, 
         {(1+z)^2\, \delta^2\over 4\,\pi\,D_L^2}\, .
\label{linefluence}
\end{equation}
The measured energies of the observed X-ray lines in the above 4 GRBs and
the Doppler factors implied by their interpretation as hydrogen
recombination features are listed in Table II. The inferred Doppler
factors are consistent with those needed in the CB model $(\delta\sim
10^3)$ to explain the intensity and duration of the GRB pulses. 

Note that the 4.4 keV line in the afterglow of GRB 991216 may be either
a hydrogen recombination edge from the same CB that produces the 3.49 keV
line, or a $\rm Ly-\alpha$ line from another CB.

\subsection{Observational Tests}

There are various independent tests 
of the  interpretation of the X-ray  lines as hydrogen recombination 
features that are Doppler shifted to X-ray energies by the CB motion:

\noindent
{\bf a. Time and duration of line emission}: 
The mean time for radiative recombination  in hydrogenic plasma is  
$\rm r_{rec}\approx 3\times 10^{10}\, T^{-1/2}\,n_e \, s^{-1}\,.$
In the observer frame this recombination time is
\begin{equation}
\rm \Delta t_L\approx {3\times 10^{10}\, (1+z)\, T^{1/2}\over
             n_e\,\delta}\, s\,.  \label{tline} \end{equation} The
emission rate of bremsstrahlung by an hydrogenic plasma is $\rm L\simeq
1.43\times 10^{-27}\, n_e^2\, T^{1/2}\, erg\, cm^{-3}\, s^{-1}\, ,$ and
the cooling time of the CB to temperature T in the observer frame is
\begin{equation} \rm t_{brem}\simeq {2.9\times 10^{11}\,(1+z)\,
T^{1/2}\over n_e\,\delta}\,s\,.  \label{tbrem} \end{equation} Thus, the
ratio $\rm\Delta t_L/t_{brem}\simeq 0.10$, where the dependence on Doppler
factor, electron density, temperature and redshift of the CB has been
canceled out, is a universal ratio for CB afterglows, independent of
their detailed properties. This prediction is consistent with the
observations of the X-ray line in GRB 980828 where $\rm\Delta
t_L/t_{brem}\simeq 0.05$ with a large ($\sim$ factor 2) uncertainty
(Yoshida et al. 1999). 

However, the above estimate is valid only for a single CB (originally
ejected, or one that was formed by overtaking and merger of separately
ejected CBs). Generally, the afterglows of different CBs cannot be
resolved either spatially or temporally. Their individual afterglows are
blended into a single afterglow. Because of their different Lorentz and
Doppler factors, X-ray line emission may extend over a much longer time,
$\rm\Delta t_L\simeq t_{brem}$.  This may be the case in the observations
of X-ray line emission during the afterglow of GRB 970508 and GRB 000214
by BeppoSAX and of GRB 991216 by Chandra which were too short both in time
and of statistics to measure accurately enough the duration of the line
emission.

\noindent
{\bf b. Photon fluences :}
Table II also reports the total photon fluence in  X-ray lines   
during the observation time and the total baryon number required
to produce the lines as predicted by  Eq.~(\ref{linefluence}).
These baryon numbers are within the range 
expected for SNe CBs/jets ($\rm N_b\simeq 6.5\times 10^{51\pm 1}$).
However, because the observation times may not have extended over the full 
time of line emission, these  baryon numbers may  underestimate
the baryon number of the CB/jet.  

\noindent 
{\bf c. Line width :} Thermal broadening of both the
recombination lines and the recombination edge are rather small. However,
the Doppler factors of the CBs decrease with time during the line emission
because of the decrease of
their Lorentz factors due to their 
deceleration by the ISM.
At late time, $\rm \gamma \sim t^{-1/3}\, ,$ for a
single CB and as a result the line energy shifts by $\rm \Delta E_L \simeq
(\Delta t_L/3\,t_{berm})\, E_L \simeq 0.04\, E_L\,$ during the line
emission. The decrease in the line energy with time during the line
emission is a clear fingerprint of the origin of the lines. When
integrated over time (in order to increase statistics) the line shift
will appear as a line broadering. The above estimate of this line 
broadening is consistent with the reported widths of the X-ray lines.

\noindent 
{\bf d. The CB radius during line emission:} The electron density
in the CB during the line emission can be inferred from the observed time
and duration of the line emission, using Eqs.~(\ref{tline}),(\ref{tbrem}).
Then, the CB radius can be estimated from the total baryon number which is
inferred from the measured line-fluence. This radius estimate is not
sensitive to the accuracy of the observations since it depends on the
third root of $\rm N_b/n_e$. The three GRBs with measured redshifts yield
$\rm R_{CB}\sim (1-2)\times 10^{15}\, cm $ for $\rm t\sim 1-2\, days$ in
the observer frame. If a CB continues to expand within the first few weeks
with the same mean speed of expansion as in the first day or two, its
radius after a month reaches $\rm \simeq (3-6)\times 10^{16}\, cm$ (only
then the swept up ISM mass begins to be comparable to the CB mass).
Indeed, from VLA observations of scintillations (Goodman 1997) in the
radio afterglow of GRB 970508 and their disappearance after a month it was
inferred (Taylor et al. 1997) that the linear size of its source a month
after burst was $\rm \approx 10^{17}\, cm\, ,$ i.e., corresponding to $\rm
R_{CB}\sim 5\times 10^{16}\, cm\,. $

\section{Conclusions}

GRBs and their afterglows may be produced
by jets of extremely relativistic cannonballs in SN 
explosions. The cannonballs which exit the supernova
shell/ejecta reheated by their collision with it, emit highly
forward-collimated radiation which is Doppler shifted to $\gamma$-ray
energy. Each cannonball corresponds to an individual pulse in a GRB. They
decelerate by sweeping up the ionised interstellar matter in front of
them, part of which is accelerated to cosmic-ray energies and emits
synchrotron radiation: the afterglow. When the cannonballs cool below 
4500K, electron-proton recombination to hydrogen produces Ly-$\alpha$
emission which is Doppler shifted to X-ray energy.      

The Cannonball Model cannot predict
the timing sequence of the GRB pulses, it fares very well in describing
the total energy, energy spectrum, and time-dependence of the individual
$\gamma$-ray pulses and their afterglow and explains the X-ray lines
observed in some GRB afterglows.

For the GRB to be observable, the CBs must be close to
the line of sight, implying that their afterglows would appear to
move superluminally. Only one of the located GRBs (980425)
is close enough to us for this superluminal displacement to be observable
with the currently available resolution. Its afterglow may by now
be too dim to be seen. Or it may not.  
If observed, the superluminal displacement of this GRB's afterglow would
be a decisive card in favour of cannonballs, as opposed to stationary
fireballs. 

The Cannonball model predicts that GRB pulses are accompanied by short
pulses of TeV neutrinos and sub TeV $\gamma$-rays, which begin and peaka 
little earlier and are much more energetic than the GRB pulses. These high
energy emission should be visible in ground based (sub TeV photon) and
underground (neutrino) telescopes.

There are other events in which a variety of GRBs could be produced by
the CB mechanism: large mass accretion
episodes in binaries including a compact object, mergers of neutron stars
with neutron stars or black holes (Paczynski 1986, Goodman et al. 1987),
transitions of neutron stars to hyperon- or quark-stars (Dar 1999; Dar and
De R\'ujula, 2000d), etc. In each case, the ejected cannonballs would make
GRBs by hitting stellar winds or envelopes, circumstellar mass or light.
I discussed only core-collapse SN explosions, as the GRBs they would
produce, although relatively ``standard'', satisfactorily
reproduce the general properties of the heterogeneous ensemble of GRBs,
their afterglows and even their X-ray line emission. 

{\bf Acknowledgements:} The material presented in this talk is basesd 
on work done in collaboration with Alvaro De R\'ujula. The research was
supported in part by the Fund for Promotion of Research at the Technion
and by the Hellen Asher Fund for Space Research. 

{}

\newpage
\vskip 0.3 true cm
{\bf 
\noindent
Table I - Gamma ray bursts of known redshift z}
\begin{table}[h]
\hspace{-.5cm} 
\begin{tabular}{|l|c|c|c|c|c|c|l|}
\hline
\hline
GRB   &z  &D$_{\rm L}$$^a$ &${\rm F_\gamma}$$^b$
&${\rm E_\gamma}^c$ & M$\; ^d$\\
\hline
970228   &0.695     &4.55  &0.17  & 0.025      & 25.2  \\
970508   &0.835     &5.70  &0.31  & 0.066      & 25.7  \\
970828   &0.957     &6.74  &7.4   & 2.06       &  ---  \\
971214   &3.418     &32.0  &1.1   & 3.06       & 25.6  \\
980425  &0.0085     &0.039 &0.44  & 8.14 E-6   & 14.3  \\
980613   &1.096     &7.98  &0.17  & 0.061      & 24.5  \\
980703   &0.966     &6.82  &3.7   & 1.05       & 22.8  \\
990123   &1.600     &12.7  &26.5  & 19.8       & 24.4  \\
990510   &1.619     &12.9  &2.3   & 1.75       & 28.5  \\ 
990712   &0.430     &2.55  & ---  & ---        & 21.8  \\
991208   &0.706     &4.64  &10.0  & 1.51       &$>$25  \\
991216   &1.020     &7.30  &25.6  & 8.07       & 24.5  \\
000131   &4.51      &      &3.51  & 11.0       &       \\
000301c  &2.040     &17.2  &2.0   & 2.32       & 27.8  \\
000418   &1.119     &8.18  &1.3   & 0.49       & 23.9  \\
000926   &2.066     &17.5  &2.5   & 2.98       & 24    \\ 
\hline
\end{tabular}
\end{table}
\vskip -0.3 true cm
\noindent
{\bf Comments:} $a$: Luminosity distance in Gpc (for $\rm \Omega_m=0.3,
\; \Omega_\Lambda=0.7$ and ${\rm H_0=65\, km\, s^{-1}\,Mpc^{-1}}$.
$b$: BATSE $\gamma$--ray fluences in units of
$10^{-5}$ erg cm$^{-2}$. $c$: (Spherical) energy in units of  $10^{53}$ ergs.
$d$: R-magnitude of the host galaxy, except for GRB 990510, for
which the V-magnitude is given.

\vskip 0.3 true cm
{\bf 
\noindent
Table II - GRB afterglows with X-ray lines}
\begin{table}[h]
\hspace{-.5cm} 
\begin{tabular}{|l|c|c|c|c|c|c|c|l|}
\hline
\hline
GRB  &  z & $\rm E_{line}$& $\rm F_{line}$& $\rm \Delta t_{obs}$ & 
$\rm \delta_{line}$ & $\rm N_b$ \\ 
\hline
970508 &0.835 & 3.4  & 1.5  &  30   & 612 & $3\times 10^{51}$ \\
970828 &0.957 & 5.04 & 0.45 &  24   & 967 & $>7\times 10^{50}$ \\         
991216 &1.020 & 3.49 & 0.39 &  12   & 691 & $>1.3\times 10^{51}$ \\         
991216 &1.020 & 4.4  & 0.47 &  12   & 800 & $>1.3\times 10^{51}$ \\         
000214 &      & 4.7  & 1.0  & 104   & $\geq$ 460 &              \\
\hline
\end{tabular}
\end{table}
\vskip -0.3 true cm
{\bf Comments:} Line energies in keV. Observation times in ks.   
All lines were assumed to be a Doppler shifted $Ly-\alpha$ lines.
The line fluences (in $\rm cm^{-2}$) and the baryon numbers 
are lower limits because of partial observation times.

\newpage

\begin{figure}
\begin{center}
\vspace*{.003cm}
\hspace*{-0cm}
\epsfig{file=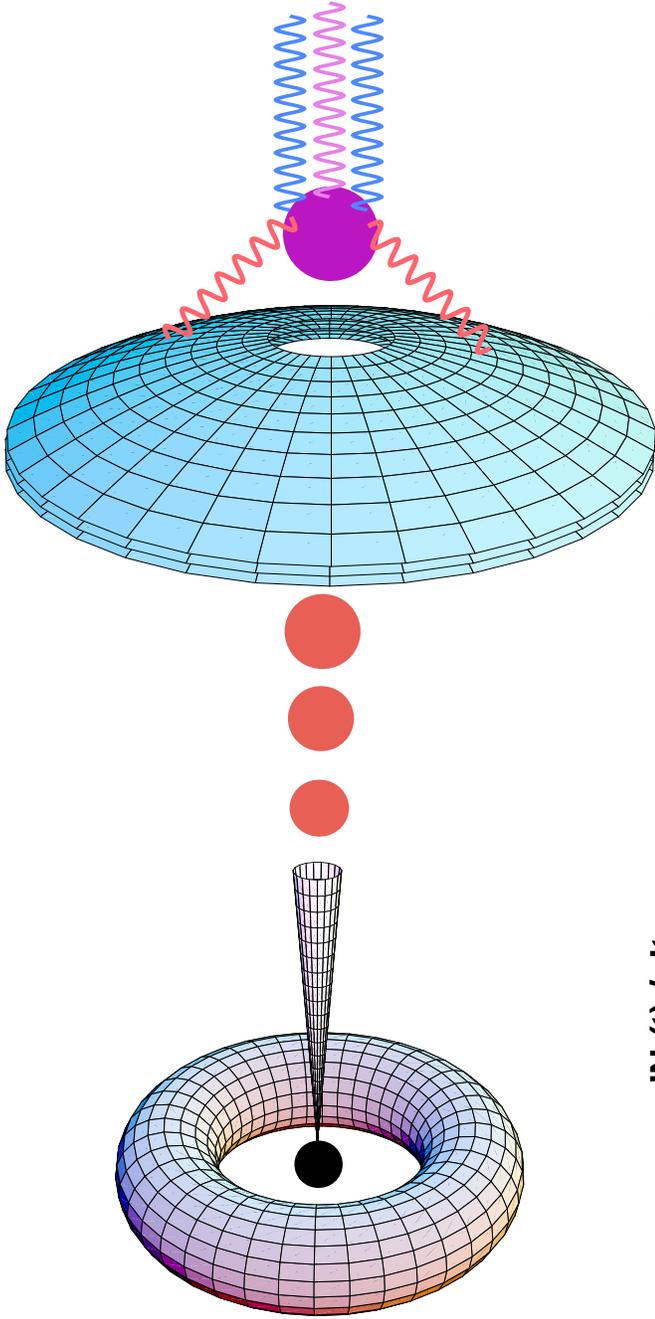,width=8.7cm}
\caption{An ``artist's view'' (not to scale) of the CB model
of GRBs and their afterglows. A core-collapse SN results in
a compact object and a fast-rotating torus of non-ejected
fallen-back material. Matter accreting (and not shown)
into the central object produces
a narrowly collimated beam of CBs, of which only some of
the ``northern'' ones are depicted. As these CBs pierce the SN shell,
they heat and reemit photons. They also scatter light from the shell.
Both emissions are Lorentz-boosted and collimated by the CBs' 
relativistic motion.}
\vspace*{-0.5cm}
\end{center}
\end{figure}

\begin{figure}
\begin{center}
\vspace*{1.0cm}
\hspace*{-1cm}
\epsfig{file=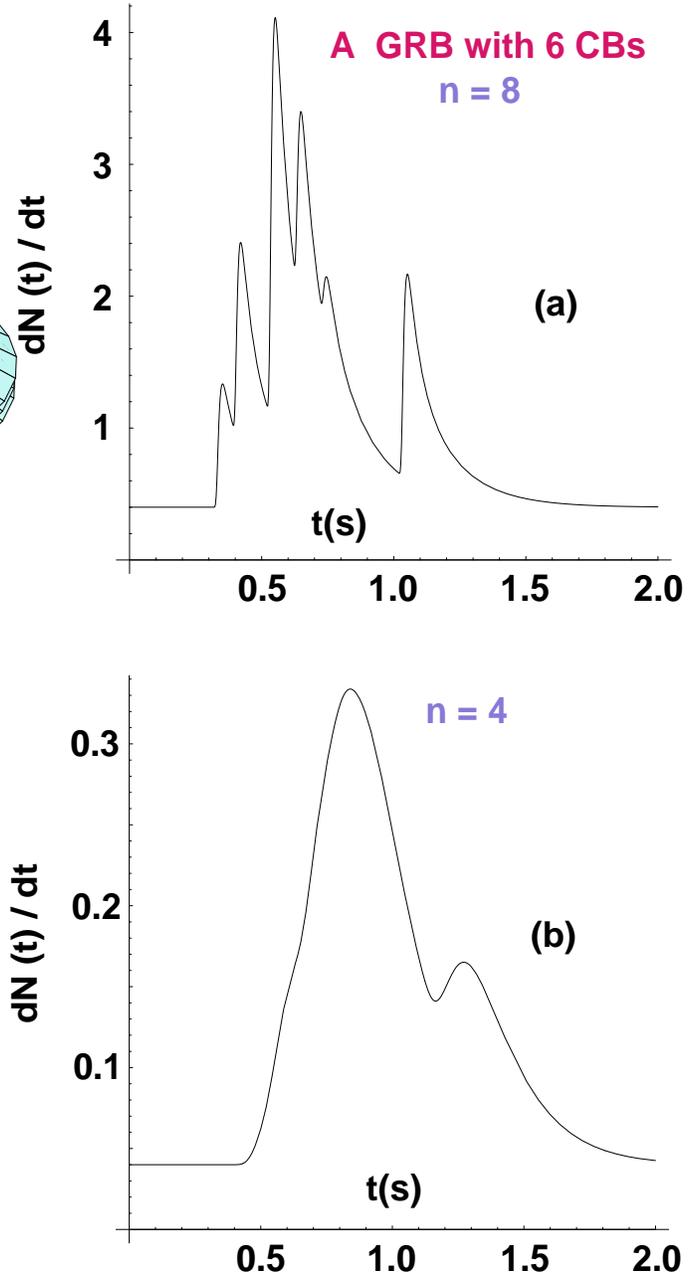,width=9cm}
\caption{ ``Synthetic'' GRB light curves, generated by shooting
six CBs at random in a 1.5 s time-interval, and with random values
of $\rm E_{CB}$ within a factor 2, 
taken from Dar and De R\'ujula 2000b.  
The only difference between (a) and (b) is that $\rm n = 8$ in (a),
while $\rm n = 4$ in (b). All other parameters in this figure have their
reference values. The figure
illustrates how a CB produces a GRB pulse, but a
GRB-pulse may not correspond to a single CB.}
\vspace*{-0.5cm}
\label{6CB}
\end{center}
\end{figure}
\begin{figure}
\vspace*{1.5cm}
\hspace*{-1cm}
\epsfig{file=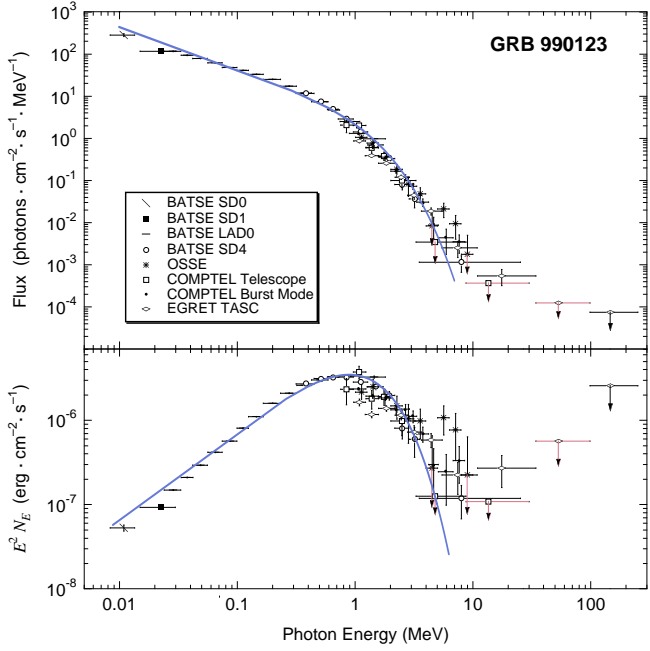,width=9cm}
\caption{Comparison of theory and observation
(taken from Dar and De R\'ujula 2000b)  
for the time-integrated energy distributions $\rm dN/dE$
and $\rm E^2\,dN/dE$, in the case of GRB 990123.
Notice that many experimental points at the
higher energies are only upper limits.}
\vspace*{-0.5cm}
\label{123}
\end{figure}
\begin{figure}
\vspace*{1.0cm}
\hspace*{-1cm}
\epsfig{file=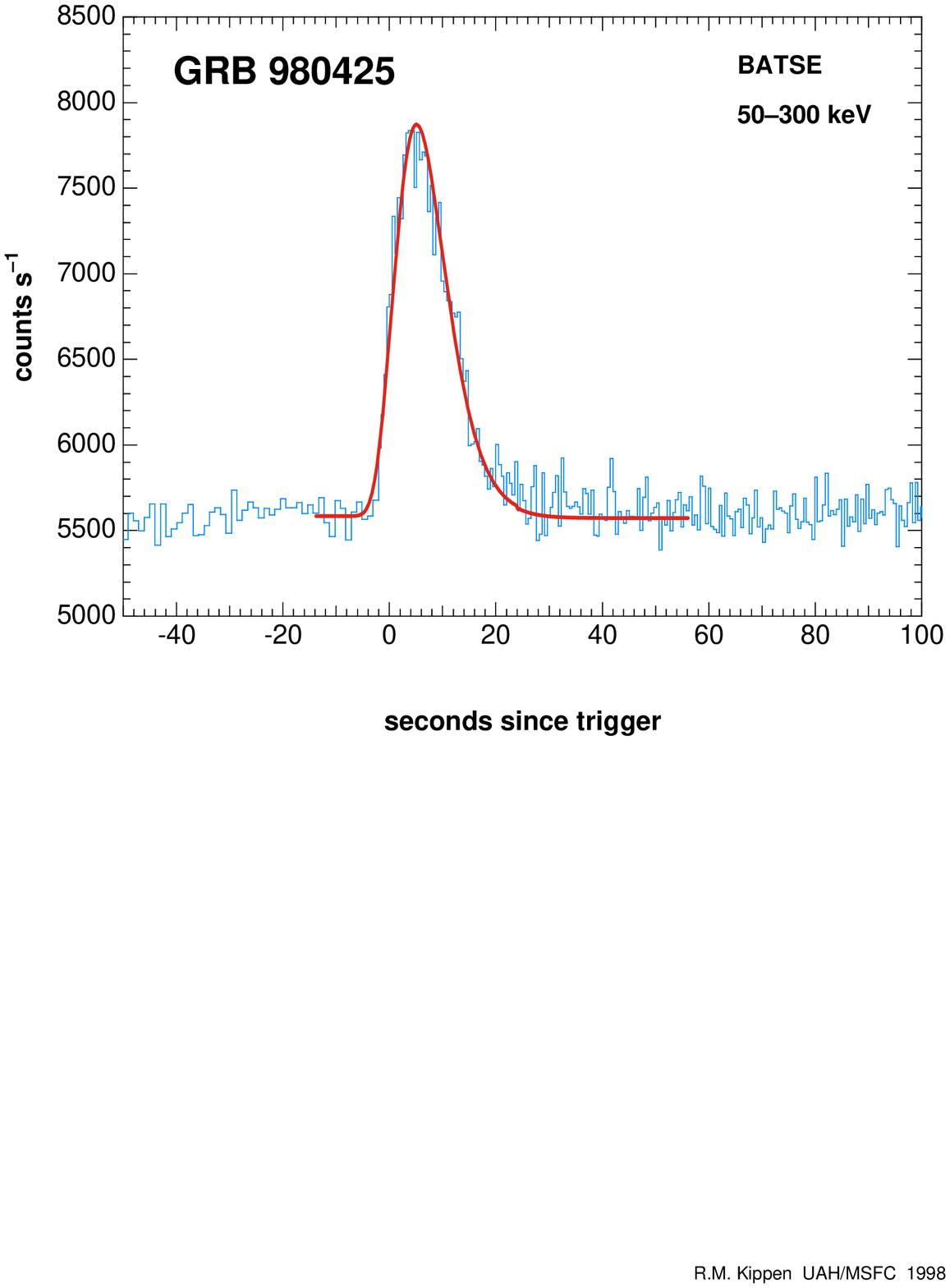,width=8cm}
\caption{Comparison of theory and observation for the
light curve of GRB 980425, in the 50-300 keV energy interval
taken from Dar and De R\'ujula 2000b.}
\vspace*{-0.5cm}
\label{425}
\end{figure}

\begin{figure}[t]
\begin{tabular}{cc}
\hskip 2truecm
\vspace*{2cm}
\hspace*{-1.7cm}
\epsfig{file=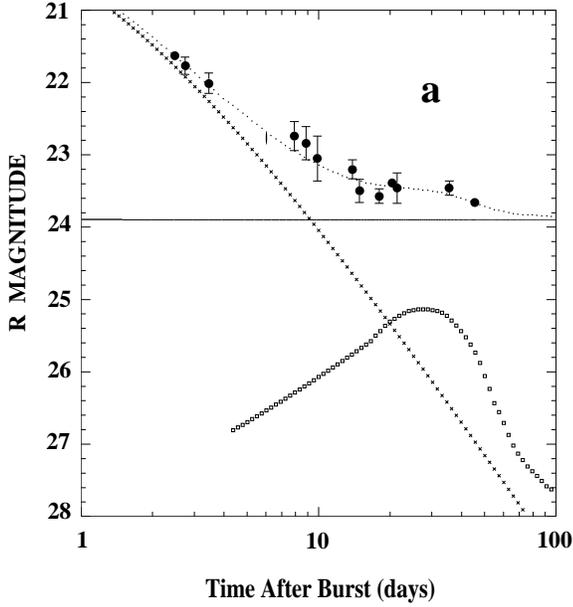,width=7.5cm} \\
\hspace*{.5cm}
\epsfig{file=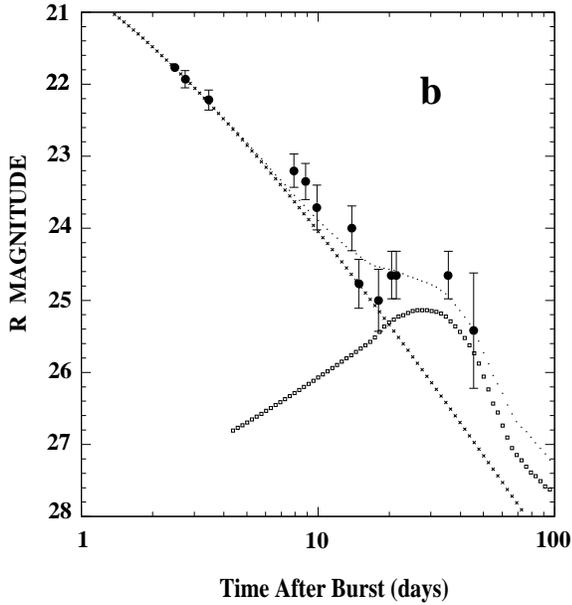,width=7.5cm}
\end{tabular}
\caption{Comparisons between the R-band light curve
for GRB 000418 (dotted lines) as calculated  
from the Cannonball model (Dar and De R\'ujula (2000a)
and the observations as 
compiled by Klose et al. (2000). a) Without subtraction of the host
galaxy's contribution: the straight line with $\rm R=23.9$ (Fruchter et al.
2000). b) With the host galaxy subtracted.
The CB's afterglow is given by Eq.(\ref{gamoft},\ref{sync})
with spectral index $\alpha=1.9$  (Klose et al. 2000) and
is indicated by crosses. The contribution
from a SN1998bw-like SN placed at z=1.11854, as in
Eq.(\ref{bw}) with
Galactic extinction ${\rm A_R=0.09}$ magnitudes,
is indicated by open squares.
The dotted line is the sum of contributions. The SN bump is
clearly discernible.}
\end{figure}

\begin{figure}
\vspace*{1cm}
\hspace*{-.7cm}
\epsfig{file=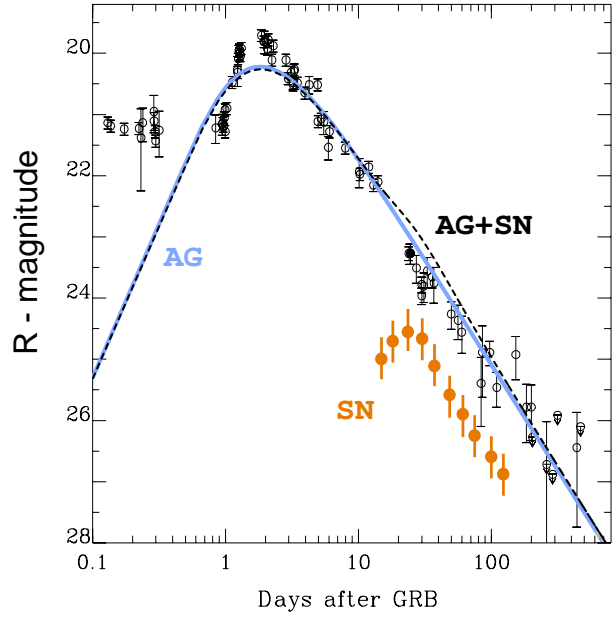,width=8cm}
\caption{The R-band light curve of the afterglow of GRB 970508 as
compiled by
Fruchter et al. (1999)
with a constant (R = 25.2 magnitude) host galaxy
subtracted from all the measurements. The blue ``AG'' curve is
given by Eqs.~(\ref{gamoft},\ref{sync}) (Dar and De R\'ujula 2000a).
The contribution from a SN1998bw-like SN,
placed at the GRB redshift $\rm z=0.835$, given by Eq.(\ref{bw}),
is indicated
(in red) and makes very little difference when added to the afterglow.}
\vspace*{-0.5cm}
\end{figure}

\begin{figure}
\vspace*{1cm}
\hspace*{-.7cm}
\epsfig{file=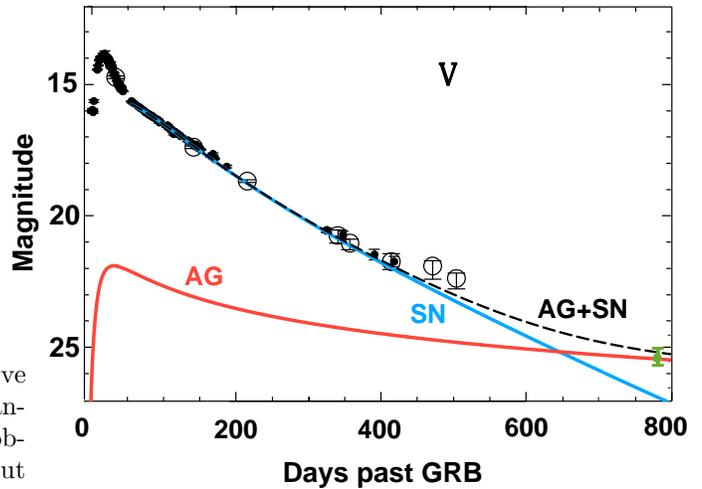,width=9cm}
\caption{The V-band light curve of
SN1998bw/GRB 980425, with
the blue ``SN'' curve a fit to the SN by Sollerman et al.~(2000).
 The red ``AG'' curve is the CB model prediction for 
the afterglow as given by 
Eqs.~(\ref{gamoft},\ref{sync}), fit
to peak at the position of the observed second radio peak, and
to reproduce the most recent observation at $\rm d=778$.
(Dar and De R\'ujula 2000a). The dashed curve is the total.}
\vspace*{-0.5cm}
\end{figure}

\end{document}